\global\long\def\vsi{\bm{\sigma}}%
\global\long\def\val{\bm{\alpha}}%
\global\long\def\vnabla{\bm{\nabla}}%
\global\long\def\calL{\mathcal{L}}%
\global\long\def\ekg{W}%
\global\long\def\vk{\bm{k}}%
\global\long\def\vp{\bm{p}}%
\global\long\def\vq{\bm{q}}%
\global\long\def\vr{\bm{r}}%
\global\long\def\order#1{\mathcal{O}\left(#1\right)}%
\global\long\def\d{\mbox{d}}%
\global\long\def\az{\alpha_{Z}}%
\documentclass{appolb}
\usepackage{graphicx}
\usepackage[latin9]{inputenc}
\usepackage{bm}
\usepackage{amsmath}

\begin{document}
\title{Negative energy states in pionic hydrogen%
\thanks{Preprint Alberta Thy 1-22. Results of this paper were partially presented at Matter to
  the Deepest, September 15-17, 2021, Institute of Physics, University
of Silesia, Poland.}
}
\author{Andrzej Czarnecki
\address{Department of Physics, University of Alberta, Edmonton,
  Alberta T6G 2E1, Canada}
}
\preprint{}
\maketitle
\begin{abstract}
Probabilities of finding an antiparticle in an atom or ion containing
a particle of spin 1/2 or spin 0 are determined. The spin 1/2 case was
previously solved by Hans Bethe and his work is summarized. The spin 0
case is treated numerically for an arbitrary atomic number and
analytically for small atomic numbers. The main  tool for the spin 0
case is the Feshbach-Villars representation of the Klein-Gordon equation.
\end{abstract}

\section{Introduction}

Solution of a Dirac equation with a Coulomb potential has a well-defined
energy, equal to the electron rest energy decreased by the binding,
which amounts  to about $13.6$
eV in case of hydrogen. However, a decomposition of the full solution of the
Dirac equation into plane waves contains both positive and negative
energy solutions of the free Dirac equation. Positive energy solutions
alone do not form a complete basis. Negative energy components,
describing antiparticles resulting from the  virtual pair production,
contribute very little to the norm of the wave function, only a fraction
of a percent even for heavy ions like the hydrogen-like lead with
the atomic number $Z=82$. For smaller $Z$, this contribution decreases
further and for small $\alpha_{Z}=Z\alpha$, where $\alpha\simeq1/137$
denotes the fine structure constant, becomes approximately $8\alpha_{Z}^{5}/(15\pi)$
\cite{Bethe:1948-H} in the ground state. Throughout this paper we
focus on the ground state only. 

Despite the smallness of their contribution to the norm, the negative
energy states have been found to contribute significantly to some
processes, on par with positive energies. For example, when a muon
bound in an atom decays, there is some probability that the resulting
electron remains bound. In this process, negative energy components of
the muon and of the electron wave functions play an important role
\cite{aslam:2019aa,Aslam:2020pqn}. This is counterintuitive.  For
example, in an earlier study of the bound muon decay, these
negative-energy contributions were neglected, which led to a
significant error \cite{Greub:1994fp}.

How large are negative energy contributions in the case of a spinless
particle like a pion, bound in a hydrogen-like atom? In the present
paper, this question is answered. This study is motivated by experiments
with pionic atoms carried out at the Paul Scherrer Institute \cite{Hori:2020aa,SoterPhD}.
We find that the probability of finding an antipion in a pionic atom
is $2\alpha_{Z}^{5}/(15\pi)$, a factor of 4 smaller than in the fermionic
case.

Section \ref{sec:negDir} reviews
Bethe's work on antiparticles in the Dirac equation. In Section \ref{sec:Pionic-atoms-and}
we summarize a two-component wave function formalism for the Klein-Gordon
equation which makes the negative energy contributions explicit. 
Probability of finding antiparticles is computed numerically  using
the momentum-space wave function (Subsection
\ref{subsec:Feshbach-Villars-representation-}). 
An analytic result is found for small $\az$ using an integral equation
for the wave function (Subsection \ref{subsec:Integral-KG-equation}).
Section \ref{sec:Conclusions} contains conclusions. An appendix reviews
solutions of the Schrödinger and the Klein-Gordon equations with a
Coulomb potential and summarizes our convention for Laguerre polynomials.

\section{Negative energy content: the case of spin 1/2}\label{sec:negDir}
\subsection{Integral Dirac equation\label{subsec:Integral-Dirac-equation}}

Let $\psi\left(\vr\right)$ be the spinor wave function of an electron
in the ground state of a hydrogen-like ion, with spin up. Define its
Fourier component $\phi\left(\vk,\tau\right)$ with spin projection
$\tau$ (with units such that $\hbar=c=1$),
\begin{equation}
\psi\left(\vr\right)=\sum_{\tau}\int\frac{\d^{3}k}{\left(2\pi\right)^{3}}\phi\left(\vk,\tau\right)u_{\tau}\left(\vk\right)e^{i\vk\cdot\vr}.\label{psiFT}
\end{equation}
$u_{\tau}$ are spatially-constant Dirac amplitudes for a free electron
normalized by $u_{\sigma}^{\dagger}u_{\tau}=\delta_{\sigma\tau}$,
\begin{align}
u_{1,2}\left(\vk\right) & =\sqrt{\frac{E+m}{2E}}\left(\begin{array}{c}
\varphi_{\pm}\\
\frac{\vsi\cdot\vk}{E+m}\varphi_{\pm}
\end{array}\right),\qquad\varphi_{+}=\left(\begin{array}{c}
1\\
0
\end{array}\right),\quad\varphi_{-}=\left(\begin{array}{c}
0\\
1
\end{array}\right),\label{posU}\\
u_{3,4}\left(\vk\right) & =\sqrt{\frac{E+m}{2E}}\left(\begin{array}{c}
\pm\frac{\vsi\cdot\vk}{E+m}\varphi_{\mp}\\
\mp\varphi_{\mp}
\end{array}\right),\quad E=\sqrt{m^{2}+k^{2}},\quad k=|\vk|.\label{negU}
\end{align}
They satisfy the Dirac equation in the following form,
\begin{align}
\left(\val\cdot\vk+\beta m\right)u_{\sigma}\left(\vk\right) & =E_{\sigma}u_{\sigma}\left(\vk\right),\label{usigma}\\
E_{1,2} & =E,\quad E_{3,4}=-E.
\end{align}
We are interested in small atomic numbers $Z$ such that $\az\ll1$.
The dominant Fourier component is $\phi\left(\vk,1\right)$. The other
positive energy component vanishes, $\phi\left(\vk,2\right)=0$, and
components with $\tau=3,4$ describe the tiny negative energy content.
All components are obtained by projection,
\begin{equation}
\phi\left(\vk,\sigma\right)=\int\d^{3}re^{-i\vk\cdot\vr}\left[u_{\sigma}^{\dagger}\psi\left(\vr\right)\right].
\end{equation}
The integral form of the Dirac equation,
\begin{equation}
\left[V\left(r\right)+\val\cdot\vk+\beta m\right]\psi=W\psi,\label{DiracToAct}
\end{equation}
is also derived with this projection. Here $V$ denotes the Coulomb
potential energy, $V\left(r\right)=-\frac{\az}{r}$, and $W$ is the
total energy, $W\simeq m-\frac{\az^{2}m}{2}$. If Eq.~\eqref{DiracToAct}
is multiplied with $\int\d^{3}re^{-i\vk\cdot\vr}u_{\sigma}^{\dagger}\left(\vk\right)$,
the first term becomes
\begin{align}
V_{\sigma}^{\prime}(\vk) & =\int\d^{3}re^{-i\vk\cdot\vr}\left[u_{\sigma}^{\dagger}\left(\vk\right)\psi\left(\vr\right)\right]V\left(\vr\right)\\
 & =\int\frac{\d^{3}q}{\left(2\pi\right)^{3}}\int\d^{3}re^{-i\left(\vk-\vq\right)\cdot\vr}\left[u_{\sigma}^{\dagger}\left(\vk\right)\psi\left(\vr\right)\right]\overbrace{\int\d^{3}r^{\prime}e^{-i\vq\cdot\vr^{\prime}}V\left(\vr^{\prime}\right)}^{V\left(\vq\right)}\\
 & =\int\frac{\d^{3}q}{\left(2\pi\right)^{3}}V\left(\vq\right)\int\d^{3}re^{-i\left(\vk-\vq\right)\cdot\vr}\left[u_{\sigma}^{\dagger}\left(\vk\right)\psi\left(\vr\right)\right].
\end{align}
Substitute $\psi$ from Eq.~\eqref{psiFT},
\begin{align}
V_{\sigma}^{\prime}(\vk) & =\int\frac{\d^{3}q}{\left(2\pi\right)^{3}}V\left(\vq\right)\sum_{\tau}\phi\left(\vk-\vq,\tau\right)\underbrace{\left[u_{\sigma}^{\dagger}\left(\vk\right)u_{\tau}\left(\vk-\vq\right)\right]}_{\left\langle \vk,\sigma|\vk-\vq,\tau\right\rangle }.
\end{align}
The conjugate of \eqref{usigma} is
$
u_{\sigma}^{\dagger}\left(\vk\right)\left(\val\cdot\vk+\beta m\right)=E_{\sigma}u_{\sigma}^{\dagger}\left(\vk\right)$,
so the last three terms of \eqref{DiracToAct} give
\begin{align}
\int\d^{3}re^{-i\vk\cdot\vr}u_{\sigma}^{\dagger}\left(\vk\right)\left(\val\cdot\vk+\beta m-W\right)\psi & =\left[E_{\sigma}-W\right]\phi\left(\vk,\sigma\right).
\end{align}
Fourier-transforming the Coulomb potential,
$\int\d^{3}r e^{-i\vq\cdot\vr}V\left(\vr\right)$ $=-\frac{4\pi\az}{q^{2}}$,
\begin{equation}
\left(W-E_{\sigma}\right)\phi\left(\vk,\sigma\right)=-4\pi\az\int\frac{\d^{3}q}{\left(2\pi\right)^{3}}\frac{1}{q^{2}}\sum_{\tau}\phi\left(\vk-\vq,\tau\right)\left\langle \vk,\sigma|\vk-\vq,\tau\right\rangle .\label{IntEq}
\end{equation}

\subsection{Solution of the Dirac equation for spin 1/2\label{subsec:NegE} }

\begin{figure}[h]
\centering
\includegraphics[scale=0.6]{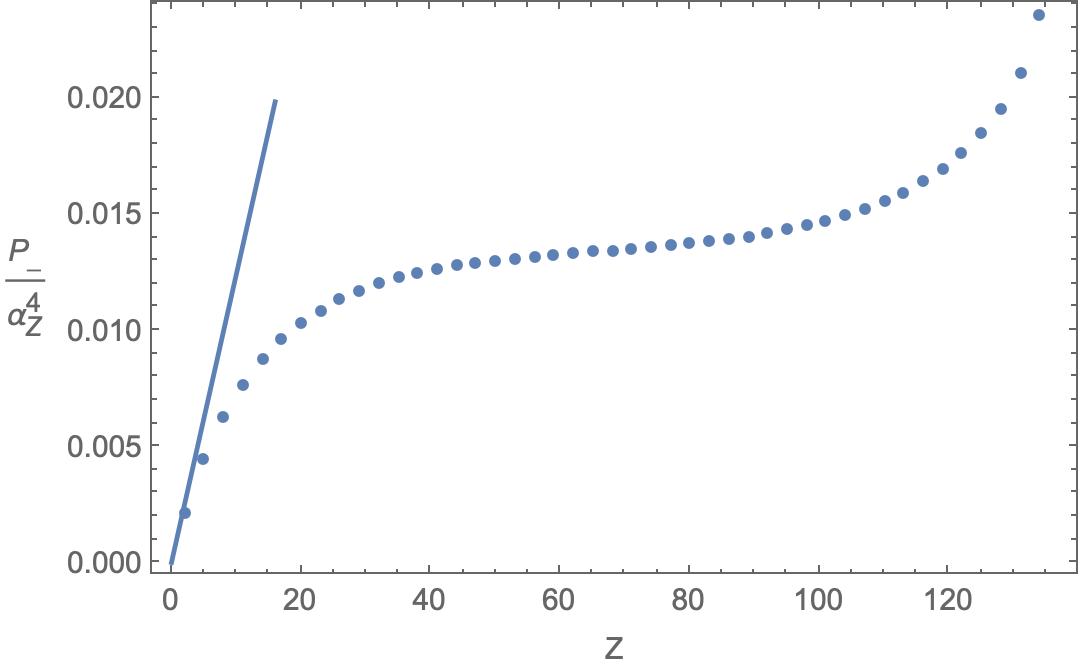}

\caption{Probability of finding spin 1/2 electrons with negative energies,
$P_{-}=\int\left(\left|\phi\left(\protect\vp,3\right)\right|^{2}+\left|\phi\left(\protect\vp,4\right)\right|^{2}\right)\frac{d^{3}p}{\left(2\pi\right)^{3}}$,
divided by $\protect\az^{4}$, as a function of the atomic number
$Z$, evaluated with $\phi\left(\protect\vp,\sigma\right)$ computed
in Ref.~\cite{Aslam:2020pqn} (dots for every third integer $Z$).
For $Z=5$ the value is about 0.004, in agreement with $\frac{8}{15\pi}\cdot5\alpha=0.006$
predicted by Eq.~\eqref{Bethe:39-1}. The first dot is for $Z=2$,
just above 0.002, in even better agreement with $\frac{8}{15\pi}\cdot2\alpha=0.0025$.
When $Z\alpha\to1$, $P_{-}$ seems to tend to 0.0329. The solid line
shows the small $Z$ behavior predicted by Eq.~\eqref{Bethe:39-1}.
\label{fig:NumericalB2}}
\end{figure}

In Eq.~\eqref{IntEq} set $W\simeq m$, $E_{\sigma}\left(\vk\right)=-\sqrt{m^{2}+k^{2}}=-E$,
and change the integration momentum $\vq\to\vp=\vk-\vq$
\begin{equation}
\left(m+E\right)\phi\left(\vk,\sigma\right)=-4\pi\az\int\frac{\d^{3}p}{\left(2\pi\right)^{3}}\frac{1}{\left(\vp-\vk\right)^{2}}\sum_{\tau}\phi\left(\vp,\tau\right)\left\langle \vk,\sigma|\vp,\tau\right\rangle .\label{Bethe:32}
\end{equation}
In the first approximation, neglect $\vp$ where possible, arguing
that it introduces higher order corrections in $\az$. Since the second
spinor component of the spin-up wave function $\psi$ vanishes, form
such a linear combination of $u_{3,4}$ in Eq. \eqref{negU} that
its second component is also zero, $u_{n}=\frac{1}{\sqrt{2E\left(E-m\right)}}\left(\begin{array}{c}
E-m\\
0\\
-k_{z}\\
-k_{+}
\end{array}\right)$. Since the wave function in the momentum space is peaked at zero
momentum, $\left(\vp-\vk\right)^{2}$ in the denominator can be approximated
by $k^{2}$ and taken out of the integral. Also, neglecting corrections
$\order{\vp/m}$, only $\sigma=n$ and $\tau=1$ contribute, $\left\langle \vk,n|0,1\right\rangle =\sqrt{\left(E-m\right)/2E}$,
\begin{align}
\phi_{-}\left(\vk\right) & \simeq-\frac{4\pi\az}{\left(m+E\right)k^{2}}\sqrt{\frac{E-m}{2E}}\int\frac{\d^{3}p}{\left(2\pi\right)^{3}}\phi\left(\vp,1\right)\label{x}\\
 & =-\frac{4\pi\az}{\left(m+E\right)k^{2}}\sqrt{\frac{E-m}{2E}}\psi\left(0\right),\label{Bethe:35-1}
\end{align}
where for the spatial wave function at the origin the non-relativistic
result can be used, $\psi\left(\vr=0\right)\simeq\sqrt{\frac{\az^{3}m^{3}}{\pi}}$.
This is the only characteristic of the wave function we need to determine
the negative energy amplitude to the leading order in $\az$. This
reflects creation of particle-antiparticle 
pairs only in the vicinity of the origin, where the potential is strong.
The resulting probability of finding negative energy states is
\begin{align}
P_-(Z) & =\int\frac{\d^{3}k}{\left(2\pi\right)^{3}}\left|\phi_{-}(\vk)\right|^{2}\\
 & =4\frac{m^{3}\az^{5}}{\pi}\int_{0}^{\infty}\d k  k^2 \left[\frac{1}{\left(m+E\right)\left(E^{2}-m^{2}\right)}\sqrt{\frac{E-m}{E}}\right]^{2}.
\end{align}
Use $k\d k=E\d E$ and change variables to $E=m\epsilon$,
\begin{align}
P_-(Z) & =4\frac{\az^{5}}{\pi}\int_1^{\infty}\frac{\d\epsilon}{\left(\epsilon+1\right)^{7/2}\left(\epsilon-1\right)^{1/2}}=\frac{8\az^{5}}{15\pi}.\label{Bethe:39-1}
\end{align}
This agrees with the numerical evaluation of $P_{-}(Z)$ presented in
Fig.~\ref{fig:NumericalB2}. Dots in that figure show $P_{-}/\az^{4}$
from a numerical integration of the negative energy components of
the exact solution of the Dirac equation with the Coulomb potential,
obtained in \cite{Aslam:2020pqn}. When $Z$ is small, these dots
come close to the straight line predicted by Eq.~\eqref{Bethe:39-1}.
However, already for $Z=8$, the straight line exceeds the numerical
value by 59 per cent, even though $\left(Z=8\right)\cdot\alpha$ is less
than $0.06$. Very likely higher-order effects in $\az$, not included
in Eq.~\eqref{Bethe:39-1}, are logarithmically enhanced. 

\section{Pionic atoms and the Klein-Gordon equation\label{sec:Pionic-atoms-and}}

We now proceed to an idealized description of a hydrogen-like ion
with the electron replaced by a negative pion $\pi^{-}$, assumed
to be point-like, stable, and not strongly interacting. 
The probability of negative energy components in its
wave function is the spin 0 analogue of Eq.~\eqref{Bethe:39-1},
which was derived for spin 1/2. We set out to derive it.

The spin 0 wave function
is described by the Klein-Gordon (KG) equation.
Decomposition of KG wave functions into plane waves with positive
and negative energies was studied by Feshbach and Villars (FV) \cite{Feshbach:1958wv}.
We shall first summarize the integral equation they derived and then
solve it with the approximation method described in Section \ref{subsec:NegE}.

\subsection{Feshbach-Villars representation of the KG wave function\label{subsec:Feshbach-Villars-representation-}}

\begin{figure}[h]
\centering
\includegraphics[scale=0.6]{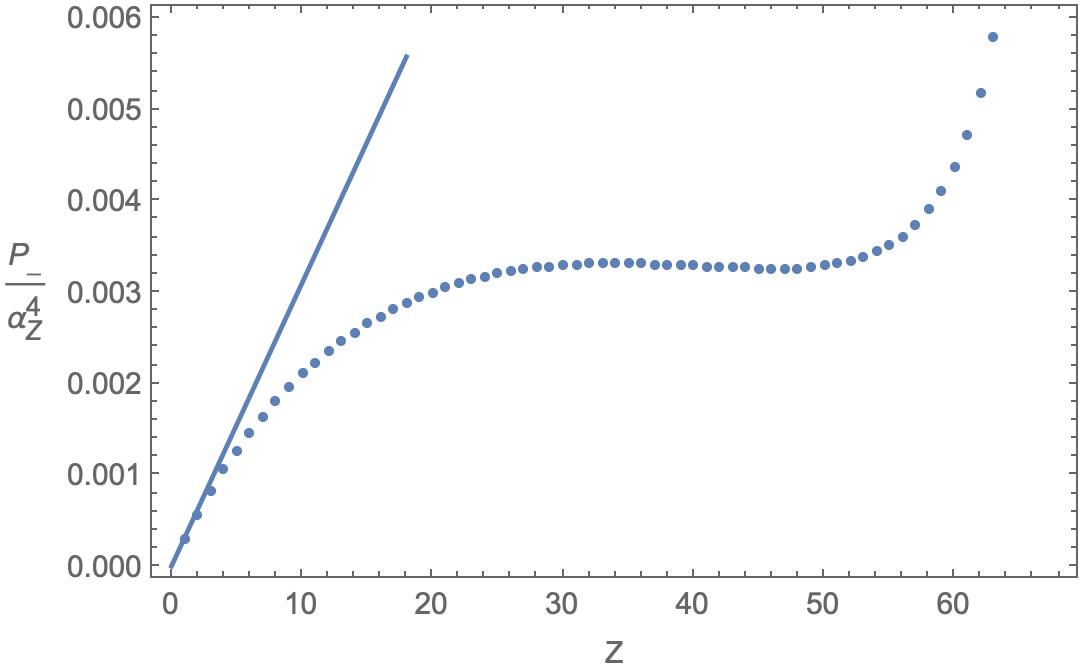}

\caption{Probability of finding spin 0 particles with negative energies (the
Klein-Gordon case), divided by $\protect\az^{4}$, as a function of
the atomic number $Z$ (dots). Note that in the KG case, the field
becomes supercritical at $\protect\az=1/2$ rather than 1 as in the
Dirac case \cite{Greiner:1985ce,Klein:1974gp}. The solid line shows
the small $Z$ behavior in Eq.~\eqref{KGsmallZ}. \label{fig:NumericalKG}}
\end{figure}

Focus on the Coulomb problem with $V\left(r\right)=-\frac{\az}{r}$
and no vector potential. The KG equation is
\begin{equation}
\left[\left(i\partial_{t}-V\right)^{2}+\nabla^{2}-m^{2}\right]\psi\left(\vr,t\right)=0.
\end{equation}
The two component wave function, which we denote with a capital letter $\Psi$, satisfying a first-order equation
in time, is
\begin{equation}
\Psi\left(\vr,t\right)=\left(\begin{array}{c}
\phi\\
\chi
\end{array}\right)=\frac{1}{\sqrt{2}m}\left(\begin{array}{c}
m+i\partial_{t}-V\\
m-i\partial_{t}+V
\end{array}\right)\psi\left(\vr,t\right).\label{KG1stOrder}
\end{equation}
The solution has the form $\Psi\left(\vr,t\right)=\Psi\left(\vr\right)e^{-iWt}$
where $W$ is the energy eigenvalue. For the Coulomb problem $W=m-\frac{m\az^{2}}{2}+\order{\az^{4}}$.
Assume that $W$ has been determined and focus on the
time-independent part of the wave function. Use such units
of energy that $m=1$. In momentum space,
\begin{equation}
\Psi\left(\vp\right)=\int\d^{3}re^{-i\vp\cdot\vr}\Psi\left(\vr\right),
\end{equation}
the wave function can be decomposed into plane waves with positive
and negative energies,
\begin{equation}
\Psi\left(\vp\right)=u\left(p\right)\Psi_{0}^{\left(+\right)}\left(\vp\right)+v\left(p\right)\Psi_{0}^{\left(-\right)}\left(\vp\right),\quad p=\left|\vp\right|,
\end{equation}
with $\Psi_{0}^{\left(\pm\right)}\left(\vp\right)$ being an orthonormal
basis, analogous to $u_{1,\dots,4}$ in Eqs.~\eqref{posU} and \eqref{negU}.
This basis diagonalizes the free-particle Hamiltonian, explicitly
decoupling positive and negative energy solutions (see Eq.~\eqref{diag}). Coefficients $u,v$ are related to $\phi,\chi$
by a unitary transformation; using $E_{p}=\sqrt{1+p^{2}}$,
\begin{equation}
\Psi^{\#}=\left(\begin{array}{c}
u\left(p\right)\\
v\left(p\right)
\end{array}\right)=U^{-1}\left(\begin{array}{c}
\phi\left(p\right)\\
\chi\left(p\right)
\end{array}\right),\, U^{-1}=\frac{1}{2\sqrt{E_{p}}}\left(\begin{array}{cc}
E_{p}+1 & E_{p}-1\\
E_{p}-1 & E_{p}+1
\end{array}\right).\label{PsiInFree}
\end{equation}
Fourier components $\phi,\chi$ can be expressed in closed form, obtained
from the configuration space wave function (see Appendix \ref{sec:Scalar-wave-equations}),
\begin{align}
\phi\left(p\right) & =\phantom{-}a\left(p\right)\left(\sqrt{p^{2}+\nu}s_{1}+\frac{1-\nu+\sqrt{1-\nu}}{\sqrt{\nu}}s_{2}\right)\\
\chi\left(p\right) & =-a\left(p\right)\left(\sqrt{p^{2}+\nu}s_{1}+\frac{1-\nu-\sqrt{1-\nu}}{\sqrt{\nu}}s_{2}\right)\\
a\left(p\right) & =\frac{2^{1-\nu}\sqrt{\pi}\sqrt[4]{\nu\left(1-\nu\right)}\Gamma\left(1-\nu\right)}{p\sqrt{\Gamma\left(2-2\nu\right)}}\left(\frac{p^{2}+\nu}{\nu}\right)^{\frac{\nu}{2}-1},\\
s_{n=1,2} & =\sin\left(\left(n-\nu\right)\arctan\frac{p}{\sqrt{\nu}}\right),\quad\nu=\frac{1}{2}-\sqrt{\frac{1}{4}-\az^{2}}.
\end{align}
In the weak field limit when $\az\ll1$, $\nu$ is of order $\az^{2}$
and so is the typical $p^{2}$. Then $\chi\ll\phi$ and $\chi$ is
analogous to the small component of the Dirac wave function. Similarly,
$v\ll u$. It is $v\left(p\right)$ that determines the probability
$P_{-}$ of finding antiparticles,
\begin{equation}
P_{-}\left(Z\right)=\int\frac{\d^{3}p}{\left(2\pi\right)^{3}}\left|v\left(p\right)\right|^{2}.
\end{equation}
The result is plotted in Fig.~\ref{fig:NumericalKG} for $Z$ up
to 68. 
Note that for larger $Z$, when $\az>1/2$, the field becomes supercritical
\cite{Greiner:1985ce,Klein:1974gp}, unlike in the Dirac equation
case which requires $\az>1$ for super-criticality. For small $\az$,
numerical results plotted in Fig.~\ref{fig:NumericalKG} indicate
the behavior
\begin{equation}
P_{-}\left(Z\to0\right)=\frac{2\az^{5}}{15\pi},\label{KGsmallZ}
\end{equation}
a four times smaller slope that in the Dirac equation case, Eq.~\eqref{Bethe:39-1}.
Eq.~\eqref{KGsmallZ} can be confirmed analytically with the help
of an integral equation, as we now proceed to demonstrate.

\subsection{Integral KG equation\label{subsec:Integral-KG-equation}}

\begin{figure}[h]
\centering\hspace*{-15mm}\includegraphics[scale=0.4]{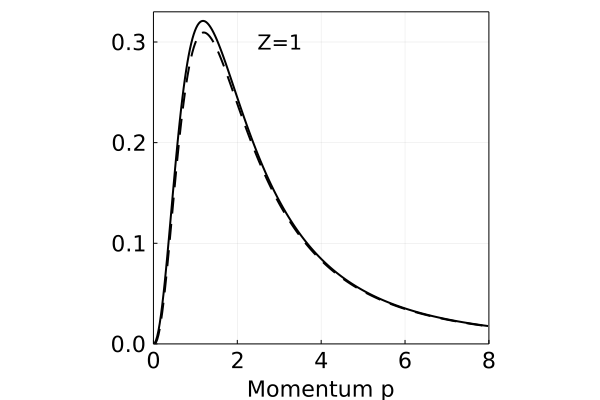}\hspace*{-20mm}\includegraphics[scale=0.4]{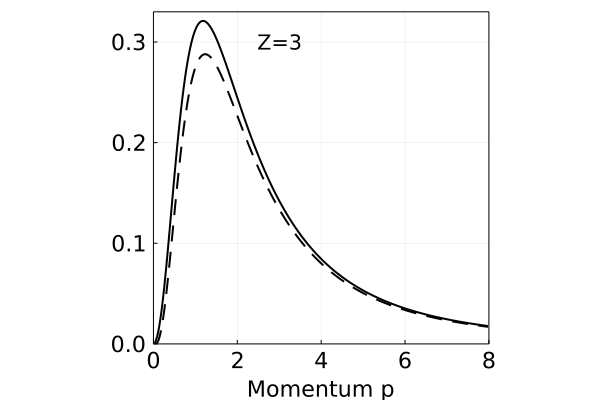}

\caption{Numerical (solid line) and approximate analytical (dashed) solutions
of the integral equation, Eq.~\eqref{FV2.55b}. The error decreases with decreasing
$Z$: neglecting $q$ under the
integral is  sound. For each $Z$, curves were rescaled
to make the area under the solid curve equal 1.\label{fig:approx} }
\end{figure}

Following Feshbach and Villars, write down the first order equation
for the wave function $\Psi^{\#}$ decomposed into free particle
solutions,
Eq.~\eqref{PsiInFree}. In momentum space, position operator is
represented by $i\vnabla_{p}$.
Using $\tau_2
= \begin{pmatrix}0&-i\\i&0\end{pmatrix}$ and $\tau_3 = \begin{pmatrix}1&0\\0&-1\end{pmatrix}$,
\begin{align}
i\partial_{t}\Psi^{\#}=U^{-1}i\partial_{t}\Psi & 
=U^{-1}\left\{ \left(\tau_{3}+i\tau_{2}\right)\frac{\vp^{2}}{2m}+\tau_{3}m+V\left(i\vnabla_{p}\right)\right\} U\Psi^{\#}.
\end{align}
With identities
\begin{align}
U^{-1}\tau_{3}U & =\frac{\left(E_{p}^{2}+1\right)\tau_{3}+\left(1-E_{p}^{2}\right)i\tau_{2}}{2E_{p}},\\
U^{-1}\tau_{2}U & =\frac{\left(E_{p}^{2}+1\right)\tau_{2}+\left(E_{p}^{2}-1\right)i\tau_{3}}{2E_{p}},
\end{align}
the free-particle part of the Hamiltonian is diagonal,
\begin{equation}
U^{-1}\left[\left(\tau_{3}+i\tau_{2}\right)\frac{\vp^{2}}{2}+\tau_{3}\right]U=E_{p}\tau_{3},
\label{diag}
\end{equation}
and the wave equation becomes
\begin{align}
i\partial_{t}\Psi^{\#} & =E_{p}\tau_{3}\Psi^{\#}+U^{-1}\left(\vp\right)V\left(i\vnabla_{p}\right)U\left(\vp\right)\Psi^{\#}\left(\vp\right).
\end{align}
For the Coulomb potential,
\begin{equation}
i\partial_{t}\Psi^{\#}=E_{p}\tau_{3}\Psi^{\#}-4\pi\az U^{-1}\left(\vp\right)\int\frac{\d^{3}q}{\left(2\pi\right)^{3}}\frac{U\left(\vq\right)\Psi^{\#}\left(\vq\right)}{\left(\vp-\vq\right)^{2}}.\label{FV2.54}
\end{equation}
We are interested in the equation for the lower component $v$. With
$i\partial_{t}\to W$ and neglecting $v$ in the right hand side since $v\ll u$, 
\begin{align}
\left(W+E_{p}\right)v & =-4\pi\az\int\frac{\d^{3}q}{\left(2\pi\right)^{3}}\frac{\left(E_{p}-E_{q}\right)u}{2\sqrt{E_{p}E_{q}}\left(\vp-\vq\right)^{2}}.\label{FV2.55b}
\end{align}
Following
the approximation discussed below Eq.~\eqref{Bethe:32}, we neglect
$q$ where possible under the integral and find
\begin{align}
v\left(p\right) & \simeq-\frac{2\pi\az\left(E_{p}-1\right)}{\sqrt{E_{p}}p^{2}\left(1+E_{p}\right)}\int\frac{\d^{3}q}{\left(2\pi\right)^{3}}u\left(q\right)\\
 & \simeq-\frac{2\pi\az\psi(0)}{\sqrt{E_{p}}\left(1+E_{p}\right)^{2}},\label{vApprox}
\end{align}
as obtained  in Ref.~\cite{PRA.51.1804}. 
To check this approximation, we plot in Fig.~\ref{fig:approx}
the numerical solution of the integral equation \eqref{FV2.55b} (solid line), and the analytical
result in Eq.~\eqref{vApprox} (dashed). As $Z$ tends to zero, the two
curves become closer. This illustrates that the momentum wave function
strongly decreases with increasing momentum; the typical momentum
is $\az$.
 The integration based on the analytical formula in Eq.~\eqref{vApprox}
is elementary,
\begin{align}
P_{-}\left(Z\to0\right) & =\int\frac{\d^{3}p}{\left(2\pi\right)^{3}}\left|v\left(p\right)\right|^{2}
 & =2\frac{\az^{5}}{\pi}\int_{1}^{\infty}\frac{\sqrt{E_{p}-1}\d E_{p}}{\left(1+E_{p}\right)^{7/2}}=\frac{2\az^{5}}{15\pi},
\end{align}
in agreement with Eq.~\eqref{KGsmallZ}.

\section{Conclusions \label{sec:Conclusions}}

We have determined the probability of finding an antiparticle
in two systems: the previously studied spin 1/2 particle in the Coulomb potential
of a point-like, static nucleus with atomic number $Z$, and an analogous system with a spin 0 particle (an
idealized pionic atom or ion). In both cases the probability is suppressed
by five powers of $Z\alpha$, and, for small $Z$, is smaller by a factor 4 in the
spin 0 case. We found that both cases, described by the Dirac and
the Klein-Gordon equations, can be treated in an analogous manner.
In the future, it would be interesting to interpret these results
in terms of Feynman diagrams. 

\section*{Acknowledgements}

I thank Alexander Penin and Vladimir Shabaev for useful discussions
and M.~Jamil Aslam and M.~Mubasher for reading the manuscript and
suggesting improvements.
This research was supported by the Natural Sciences and Engineering
Research Council of Canada (NSERC), by WestGrid (www.westgrid.ca),
and by Compute Canada (www.computecanada.ca).

\appendix

\section{Klein-Gordon equation with a Coulomb potential} \label{sec:Scalar-wave-equations}

We consider a pion in the Coulomb field of an infinitely heavy point-like
nucleus with charge $Ze$.  We first summarize the solution of the radial Schrödinger equation with
a Coulomb potential, to emphasize its similarity with the KG case,
treated in detail. The Schrödinger equation  reads (as in the main text, we use such units
that $\hbar=c=1$, but we keep $m$ explicit)
\begin{equation}
\left[\left(\partial_{r}+\frac{1}{r}\right)^{2}-\frac{l\left(l+1\right)}{r^{2}}+\frac{2\az m}{r}+2m\left(E-m\right)\right]R^{\text{Sch}}\left(r\right)=0,
\end{equation}
where $E-m=-\frac{\az^{2}m}{2\left(1+n_{r}+l\right)^{2}}$ is the
binding energy, with $n_{r}=0,1,\dots,$ denoting the radial excitation,
and $l=0,1,\dots,$ denoting the angular momentum. With the distance given
by $x=r/a$ in units of the Bohr radius $a=1/\left(\az m\right)$,  the resulting radial
wave functions \cite{Landau:1991wop} are 
\begin{equation}
R_{nl}^{\text{Sch}}\left(x\right)=-\frac{2}{n^{2}}\sqrt{\frac{\left(n+l\right)!}{\left(n-l-1\right)!}}\exp\left(-\frac{x}{n}\right)\left(\frac{2x}{n}\right)^{-l-1}L_{n+l}^{\left(-2l-1\right)}\left(\frac{2x}{n}\right).\label{R:Sch}
\end{equation}
Laguerre polynomials $L_{n}^{\left(\alpha\right)}$ are defined in
Appendix \ref{sec:AppLag}. In the ground state, $n=1$, $l=0$, the
radial wave function becomes $2\exp\left(-x\right)$. 

For the KG equation we have, from $\left[\nabla^{2}-m^{2}+\left(\ekg+\frac{\az}{r}\right)^{2}\right]\psi=0$,
\begin{equation}
\left[\left(\partial_{r}+\frac{1}{r}\right)^{2}-\frac{l\left(l+1\right)}{r^{2}}+\frac{\az^{2}}{r^{2}}+\frac{2\az\ekg}{r}+\ekg^{2}-m^{2}\right]R^{\text{KG}}\left(r\right)=0.
\end{equation} 
We rescale the distance variable, $\rho=\sqrt{m^{2}-\ekg^{2}}r$ and
replace 
\begin{equation}
l\to\lambda=\sqrt{\left(l+\frac{1}{2}\right)^{2}-\az^{2}}-\frac{1}{2},
\end{equation}
to derive the radial equation in a dimensionless form,
\begin{equation}
\left[\left(\partial_{\rho}+\frac{1}{\rho}\right)^{2}-\frac{\lambda\left(\lambda+1\right)}{\rho^{2}}+\frac{\varepsilon}{\rho}-1\right]R=0,\quad\varepsilon\equiv\frac{2\az\ekg}{\sqrt{m^{2}-\ekg^{2}}}.
\end{equation}
For large $\rho$,
\begin{equation}
\left(\rho R\right)^{\prime\prime}=\rho R\Rightarrow R\sim\frac{e^{-\rho}}{\rho},
\end{equation}
while for small $\rho$, $\partial_{\rho}^{2}\left(\rho R\right)=\frac{\lambda\left(\lambda+1\right)}{\rho^{2}}\rho R$,
so $R\sim\rho^{\lambda}$. With the substitution $R=\rho^{\lambda}e^{-\rho}L\left(\rho\right)$,
the equation for $L$ becomes
\begin{equation}
\rho L^{\prime\prime}+2\left(\lambda+1-\rho\right)L^{\prime}+\left[\varepsilon-2\left(\lambda+1\right)\right]L=0\label{EqL}
\end{equation}
Substituting a power series for $L$,
$L=\sum_{k=0}^{\infty}a_{k}\rho^{k},$ gives
a recurrence relation,
\begin{align}
\left(2\left(\lambda+1\right)+k\right)\left(k+1\right)a_{k+1}+\left[\varepsilon-2\left(\lambda+1\right)-2k\right]a_{k} & =0,
\end{align}
The series terminates if for some $k$ the coefficient of $a_{k}$
vanishes, that is when $\frac{e-2\left(\lambda+1\right)}{2}=n_{r}=0,1,\dots$.
This gives the quantization condition for the energy,
\begin{align}
\frac{\az\ekg}{\sqrt{m^{2}-\ekg^{2}}} & =1+\lambda+n_{r},
\end{align}
so that finally
\begin{equation}
\ekg=\frac{m}{\sqrt{1+\frac{\az^{2}}{\left(1+\lambda+n_{r}\right)^{2}}}}.\label{eqE}
\end{equation}
When the condition $\varepsilon-2\left(\lambda+1\right)=2n_{r}$ is
fulfilled, Eq.~\eqref{EqL} becomes
\begin{equation}
\rho L^{\prime\prime}+2\left(\lambda+1-\rho\right)L^{\prime}+2n_{r}L=0.
\end{equation}
Change the variable to $x=2\rho$ and recognize the generalized Laguerre
equation,
\begin{equation}
x\frac{\d^{2}L}{\d x^{2}}+\left(2\lambda+2-x\right)\frac{\d L}{\d x}+n_{r}L=0,
\end{equation}
whose solutions are $L\left(x\right)=L_{n_{r}}^{\left(2\lambda+1\right)}\left(x\right)$.
Remembering $\rho=\sqrt{m^{2}-\ekg^{2}}r$ we get
\begin{align}
x & =2\rho=\frac{2mr\az}{\sqrt{\left(1+\lambda+n_{r}\right)^{2}+\az^{2}}},\\
R & =Nx^{\lambda}e^{-x/2}L_{n_{r}}^{\left(2\lambda+1\right)}\left(x\right).
\end{align}
The normalization $N$ is often defined by the condition (but see
the discussion below Eq.~\eqref{R00})
\begin{align}
1 & =N^{2}\int_{0}^{\infty}x^{2\lambda}e^{-x}\left[L_{n_{r}}^{\left(2\lambda+1\right)}\left(x\right)\right]^{2}r^{2}\d r\\
 & =N^{2}s^{3}\cdot2\left(n_{r}+\lambda+1\right){n_{r}+2\lambda+1 \choose n_{r}}\Gamma\left(2\lambda+2\right).
\end{align}
In summary, the solution of the Klein-Gordon equation with the Coulomb
potential is (see also \cite{bagrov1990exact})
\begin{align}
R_{n_{r}l} & =\left(\frac{\az m}{\sqrt{\left(1+\lambda+n_{r}\right)^{2}+\az^{2}}}\right)^{3/2}\frac{2\sqrt{n_{r}!}x^{\lambda}e^{-x/2}L_{n_{r}}^{\left(2\lambda+1\right)}\left(x\right)}{\sqrt{\left(1+\lambda+n_{r}\right)\Gamma\left(2+2\lambda+n_{r}\right)}},\\
x & =\frac{2\az mr}{\sqrt{\left(1+\lambda+n_{r}\right)^{2}+\az^{2}}},\qquad\lambda=\sqrt{\left(l+\frac{1}{2}\right)^{2}-\az^{2}}-\frac{1}{2},\\
\ekg & =\frac{m}{\sqrt{1+\frac{\az^{2}}{\left(1+\lambda+n_{r}\right)^{2}}}}.
\end{align}
Here $n_{r}$ is the degree of the radial excitation and $l$ is the
orbital quantum number. The ground state corresponds to $n_{r}=l=0$,
thus $\lambda\to\sqrt{\frac{1}{4}-\az^{2}}-\frac{1}{2}<0$. It is
convenient to introduce a positive parameter $\nu=\frac{1}{2}-\sqrt{\frac{1}{4}-\az^{2}}$
and use $\sqrt{\left(1-\nu\right)^{2}+\az^{2}}=\sqrt{1-\nu}$
\begin{align}
R_{00}\left(r\right) & =\left(2\sqrt{\nu}m\right)^{3/2-\nu}\frac{r^{-\nu}e^{-\sqrt{\nu}mr}}{\sqrt{\Gamma\left(3-2\nu\right)}}.\label{R00}
\end{align}
Return now to the issue of normalization. It is convenient to define
such $\psi\left(\vr\right)$ that $2\left[W-V\left(r\right)\right]\left|\psi\left(\vr\right)\right|^{2}$
is interpreted as charge density (with the charge of the negative
pion taken as the unit, $2\int\d^{3}r\left[W-V\left(r\right)\right]\left|\psi\left(\vr\right)\right|^{2}=1$).
To this end, in case of the ground state, include the spherical harmonic
$Y_{00}\left(\theta,\phi\right)=1/\sqrt{4\pi}$ and define \cite{Greiner:1990tz}
\begin{equation}
\psi\left(\vr\right)=\frac{\left(1-\nu\right)^{1/4}}{\sqrt{8\pi}}R_{00}\left(r\right).
\end{equation}
In Eq.~\eqref{KG1stOrder}, $\psi\left(\vr,t\right)$ equals $\psi\left(\vr\right)e^{-iWt}$
with $W=\sqrt{1-\nu}m$. 

\section{Generalized Laguerre functions: conventions}\label{sec:AppLag}

We use Laguerre functions $L_{n}^{\left(\alpha\right)}$ according
to the convention of Ref.~\cite{NISTHandbook,SchwingerQM}, which
differs from Landau and Lifshitz \cite{Landau:1991wop}, whose functions
we denote by $\calL_{n}^{m}$. Here we explain the connection between
them. We use Rodrigues formula in the form
\begin{equation}
L_{n}^{\left(\alpha\right)}\left(z\right)=\frac{e^{z}z^{-\alpha}}{n!}\frac{\d^{n}}{\d z^{n}}\left(e^{-z}z^{n+\alpha}\right),
\end{equation}
while Landau and Lifshitz use
\begin{equation}
\calL_{n}^{m}\left(z\right)=\frac{n!e^{z}}{\left(n-m\right)!}\frac{\d^{n}}{\d z^{n}}\left(e^{-z}z^{n-m}\right).
\end{equation}
Therefore,
\begin{equation}
\calL_{n}^{m}\left(z\right)=\frac{\left(n!\right)^{2}z^{-m}}{\left(n-m\right)!}L_{n}^{\left(-m\right)}\left(z\right).
\end{equation}

\newpage

\begin{thebibliography}{10}
\providecommand{\url}[1]{\texttt{#1}}
\providecommand{\urlprefix}{URL }
\providecommand{\eprint}[2][]{\url{#2}}

\bibitem{Bethe:1948-H}
H.~A. Bethe, \emph{Bemerkungen {\"uber} die Wasserstoff-Eigenfunktionen in der
  Diracschen Theorie}, Zeitschrift f{\"u}r Naturforschung A \textbf{3}, 470 --
  477 (1948).

\bibitem{aslam:2019aa}
M.~J. Aslam, A.~Czarnecki, A.~Morozova, and G.~P. Zhang, \emph{Decay of a bound
  muon into a bound electron}, Proceedings of Science \textbf{367}, 148 (2019),
  https://pos.sissa.it/367/148/.

\bibitem{Aslam:2020pqn}
M.~J. Aslam, A.~Czarnecki, G.~Zhang, and A.~Morozova, \emph{{Decay of a bound
  muon into a bound electron}}, Phys. Rev. D \textbf{102}, 073001 (2020),
  \eprint{2005.07276}.

\bibitem{Greub:1994fp}
C.~Greub, D.~Wyler, S.~Brodsky, and C.~Munger, \emph{{Atomic alchemy: Weak
  decays of muonic and pionic atoms into other atoms}}, Phys.~Rev.
  \textbf{D52}, 4028--4037 (1995), \eprint{hep-ph/9405230}.

\bibitem{Hori:2020aa}
M.~Hori, H.~Aghai-Khozani, A.~S{\'o}t{\'e}r, A.~Dax, and D.~Barna, \emph{Laser
  spectroscopy of pionic helium atoms}, Nature \textbf{581}, 37--41 (2020).

\bibitem{SoterPhD}
A.~S{\'o}t{\'e}r, \emph{Laser spectroscopy of antiprotonic and pionic helium
  atoms}, Ph.D. thesis, Ludwig-Maximilians-Universit{\"a}t M{\"u}nchen (2016).

\bibitem{Feshbach:1958wv}
H.~Feshbach and F.~Villars, \emph{{Elementary relativistic wave mechanics of
  spin 0 and spin 1/2 particles}}, Rev. Mod. Phys. \textbf{30}, 24--45 (1958).

\bibitem{Greiner:1985ce}
W.~Greiner, B.~Muller, and J.~Rafelski, \emph{{Quantum Electrodynamics Of
  Strong Fields}}, Springer, Berlin (1985).

\bibitem{Klein:1974gp}
A.~Klein and J.~Rafelski, \emph{{Bose Condensation in Supercritical External
  Fields}}, Phys. Rev. D \textbf{11}, 300 (1976).

\bibitem{PRA.51.1804}
M.~Horbatsch and D.~V. Shapoval, \emph{Analysis of the Klein-Gordon Coulomb
  problem in the Feshbach-Villars representation}, Phys. Rev. A \textbf{51},
  1804--1807 (1995).

\bibitem{Landau:1991wop}
L.~D. Landau and E.~M. Lifshits, \emph{{Quantum Mechanics}: {Non-Relativistic
  Theory}}, Butterworth-Heinemann, Oxford (1991).

\bibitem{bagrov1990exact}
V.~Bagrov and D.~Gitman, \emph{Exact Solutions of Relativistic Wave Equations},
  Springer (1990).

\bibitem{Greiner:1990tz}
W.~Greiner, \emph{Relativistic Quantum Mechanics. Wave Equations}, Springer,
  Berlin, 3rd edition (2000).

\bibitem{NISTHandbook}
F.~W.~J. Olver, D.~W. Lozier, R.~F. Boisvert, and C.~W. Clark, editors,
  \emph{NIST Handbook of Mathematical Functions}, Cambridge University Press,
  Cambridge (2011).

\bibitem{SchwingerQM}
J.~Schwinger, \emph{Quantum Mechanics: symbolism of atomic measurements},
  Springer, Berlin (2001), edited by B.-G.~Englert.

\end{thebibliography}

\end{document}